\documentclass[11pt,eqs]{article}
\usepackage{latexsym}
\usepackage{amsmath}
\usepackage{MnSymbol}
\usepackage{tikz}
\usepackage{pgfplots}
\usepackage{pgfarrows}
\usetikzlibrary{positioning}
\usepackage{hyperref}
\hypersetup{
	colorlinks=true,
	linkcolor=cyan,
	filecolor=magenta,      
	urlcolor=cyan,
}

\makeatletter
\newcommand\xleftrightarrow[2][]{%
	\ext@arrow 9999{\longleftrightarrowfill@}{#1}{#2}}
\newcommand\longleftrightarrowfill@{%
	\arrowfill@\leftarrow\relbar\rightarrow}
\makeatother

\title{Combined fermionic and bosonic T-duality of type II superstring theory with coordinate dependent RR field
	\thanks{Work supported in part by the Serbian Ministry of Education,Science and Technological Development.}}

\author{B. Nikoli\'c and D. Obri\'c \thanks{email:bnikolic, dobric}\\{\it Institute of Physics Belgrade, University of Belgrade, Pregrevica 118, Serbia} }

\begin{document}
	\maketitle
	
	\begin{abstract}
		We investigate effects of fermionic T-duality on type II superstring in presence of Ramond-Ramond (RR) field that has  infinitesimal linear dependence on bosonic coordinate $x^\mu$. Other fields are assumed to be constant. Procedure that we employ for obtaining fermionic T-dual theory is Busher procedure, where we will consider two distinct cases. One, where action has not been T-dualized along bosonic coordinates and other where it has. By analyzing these two cases, their actions and T-dual transformation laws, we obtain some insight into how background fields transform and what are necessary ingredients for emergence of fermionic non-commutativity.

		%We Investigate effects of fermionic T-duality on type II superstring that propagates in coordinate dependent Ramond-Ramond (RR) field. RR field only depend on bosonic coordinates $x^\mu$ and this dependence is linear and infinitesimal. To obtain fermionic T-duality  we will utilize Busher procedure. Procedure will be carried out in two separate cases. One, where action hasn't been T-dualized along bosonic coordinates and other where it has. By analyzing these two cases, their actions and T-dual transformation laws, we obtain some insight into how background fields transform and what are necessary ingredients for emergence of fermionic non-commutativity. 
	\end{abstract}
	
	\section{Introduction}
	\setcounter{equation}{0}
	
	%T-duality represents a map that connects different superstring theories, mapping geometry and topology from one theory onto other. This symmetry leaves partition function invariant.
	
	%While there are many different ways to represent T-duality, one of the most useful is the Busher procedure.  

	T-duality represents a map that connects different superstring theories, mapping geometry and topology from one theory to another \cite{book1}. This symmetry was originally developed with bosonic coordinates in mind, where two theories are connected by transformation laws that establish a link between coordinates \cite{bosonic t-duality historical paper}. It was not until 2008 that it has been noticed that same duality can emerge in case of fermionic coordinates. In their paper \cite{Fermionic T-duality} Berkovits and Maldacena showed that tree level superstring theories in presence of supersymmetric background fields posses new kind od symmetry. Symmetry that maps supersymmetric background fields of one theory to supersymmetric backgrounds of other theory, where dilaton and RR fields are now different. Just like in case of bosonic T-duality, mathematical machinery for obtaining T-dual theories is Busher procedure \cite{T-duality procedure 1} \cite{T-duality procedure 2} applied to fermionic coordinates $\theta^\alpha$ and $\bar{ \theta }^\alpha$. 
	
	Busher T-dualization procedure including its extension to fermionic coordinates\cite{Fermionic T-duality 2}\cite{Fermionic T-duality 3} \cite{Fermionic T-duality 4} \cite{Fermionic T-duality procedure} and its generalizations \cite{Generalized Buscher procedure}\cite{Gereralized Buscher procedure 2}\cite{Generalized Buscher procedure 3} mainly follow the same steps. We notice some global symmetry in the theory, usually shift symmetry, which is then localized by replacing partial derivatives with covariant ones. Covariant derivatives come with new gauge fields that insert new degrees of freedom into the theory, these degrees of freedom are eliminated with help of Lagrange multipliers. Next step is utilizing gauge freedom to fix starting coordinates. After that, finding equations of motion for gauge fields and inserting their solutions into the action we obtain T-dual theory. Extension of procedure for fermionic coordinates does not introduce any new steps into the play. However, extending the procedure to include coordinate dependent background fields does introduce one additional step. Namely, we need to replace all coordinates with invariant ones constructed as integrals of covariant derivatives. This step is necessary in order to preserve local shift symmetry.
	
	In our previous paper \cite{our paper 3} we demonstrated that T-dual of type II superstring which is moving in coordinate dependent RR field possesses non-commutative Poisson brackets. Since T-duality was performed only along bosonic coordinates it produced non-commutativity only between bosonic T-dual coordinates. In addition to this, we had that background fields that were constants in original theory became functions of both bosonic and fermionic coordinates in dual theory. This has left us with one open question: Would fermionic T-duality of starting theory or even theory that has been dualized along bosonic coordinates produce non-commutative relations between fermionic coordinates? While it has been shown that, in case of closed bosonic string, non-commutativity arises only in coordinates that had appeared in background fields of starting theory \cite{Integration of Poisson brackets 2}, it is not clear if that is the case for fermionic coordinates, especially when we have emergence of new coordinate dependence in background fields after bosonic T-duality.
	
	In this article, our goal is to find what effects fermionic T-duality has on action where RR field has dependence on bosonic coordinates and do these effects change for fully dualized action. By obtaining background fields in different stages of T-duality we can determine how geometry of theory changes and when the theory makes the switch from being local to non-local one. At the end we provide few notes on how fermionic T-duality interacts with bosonic T-duality in providing new non-commutative relations.

	\section{Type II superstring, choice of fields and bosonic T-duality}
	
	In this section we will present action for type II superstring in pure spinor formulation. We will also define background fields in which string propagates. Finally, we present action that has been T-dualized along bosonic coordinates.

	\subsection{Type II superstring in pure spinor formulation}
	
	The most general form of type II superstring action in pure spinor formalism \cite{Vertex operators} \cite{pure spinor formalism papers 1} \cite{pure spinor formalism papers 2} \cite{pure spinor formalism papers 3} \cite{pure spinor formalism papers 4} is given as
	\begin{equation}\label{eq:dejstvo}
		S=S_0 + V_{SG}.
	\end{equation}
	First term is action for string that propagates in flat background fields.
	\begin{equation} \label{S_0 action}
		S_0 = \int_{\Sigma}d^2 \xi \left( \frac{k}{2} \eta_{ \mu \nu } \partial_m x^\mu \partial_n x^\nu \eta^{m n} - \pi_\alpha \partial_- \theta^\alpha + \partial_+ \bar{\theta}^\alpha \bar{\pi}_\alpha \right) 
		+S_\lambda +S_{ \bar{\lambda} },
	\end{equation}
	where terms $ S_\lambda $ and $S_{\lambda}$ represent actions that are composed of pure spinors and their canonical momenta. The pure spinors satisfy pure spinor constraints
	\begin{equation}
		\lambda^\alpha (\Gamma^\mu)_{\alpha \beta} \lambda^\beta =  \bar{\lambda} (\Gamma^\mu)_{\alpha \beta} \bar{\lambda}^\beta = 0 .
	\end{equation}
	
	All modifications to flat background fields are accomplished by introducing second term in equation (\ref{eq:dejstvo}). This term is an integrated vertex operator for massless type II supergravity 
	\begin{equation}
		V_{SG} =  \int_{\Sigma}d^2 \xi (X^T)^M A_{MN} \bar{X}^N.
	\end{equation}
	In general case matrix $A_{MN}$ is composed of physical fields, their curvatures (field strengths) and auxiliary fields that can be expressed with physical ones. These fields are some functions of both bosonic coordinates $x^\mu$ and fermionic coordinates $\theta^\alpha$ and $\bar{\theta}^\alpha$. Dependence of fields on fermionic coordinates is given as expansion in powers of $\theta^\alpha$ and $\bar{\theta}^\alpha$. In our particular case, we will set all background fields except RR field to be constant. Further more, in order to simplify calculations, all terms that are non-linear in fermionic coordinates $\theta^\alpha$ and $\bar{\theta}^\alpha$ will be neglected. With these assumptions in mind we have that vectors $X^M$ and $\bar{X}^M$ and matrix $A_{MN}$ have following form
	\begin{equation}
		X^M = \left(\begin{matrix}
			\partial_+ \theta^\alpha\\
			\partial_+ x^\mu\\
			\pi_\alpha\\
			\frac{1}{2} N_+^{\mu \nu}
		\end{matrix} \right), 
		\  \bar{X}^M = \left( \begin{matrix}
			\partial_- \bar{\theta}^\lambda\\
			\partial_- x^\mu\\
			\bar{\pi}_\lambda\\
			\frac{1}{2} \bar{N}_-^{\mu \nu}
		\end{matrix} \right), \ 
		A_{MN} = \begin{bmatrix}
			0 & 0 & 0 & 0 \\
			0 & k(\frac{1}{2}g_{\mu \nu} + B_{\mu \nu}) & \bar{\Psi}_\mu^\beta & 0\\
			0 & -\Psi_\nu^\alpha & \frac{2}{k}(f^{\alpha\beta} + C_\rho^{\alpha\beta} x^\rho)& 0\\
			0 & 0 & 0 & 0
		\end{bmatrix}.
	\end{equation} 
	Pure spinor contribution to vectors $X^M$ and $\bar{X}^M$ are encoded in 
	
	\begin{equation}
		N_+^{\mu \nu} = \frac{1}{2} \omega_\alpha {( \Gamma^{[\mu\nu]}
			)^\alpha}_\beta \lambda ^\beta, \quad \bar{N}_-^{\mu\nu} = \frac{1}{2} \bar{\omega}_\alpha {(  \Gamma^{ [\mu\nu ]  }     )^\alpha}_\beta \bar{\lambda}^\beta.
	\end{equation}
	Since this term does not contribute to the vertex operator, we have that pure spinor actions are decoupled from the rest. This allows us to neglect pure spinor parts from now on.

	Our choice of matrix $A_{MN}$ is composed of following fields:  symmetric tensor $g_{\mu \nu}$, Kalb-Ramon antisymmetric tensor $B_{\mu \nu}$, Mayorana-Weyl gravitino fields $\Psi_\mu^\alpha$ and $\bar{\Psi}_\mu^\alpha$, Ramond-Ramond field $\frac{2}{k}(f^{\alpha\beta} + C_\rho^{\alpha\beta} x^\rho)$ where $f^{\alpha \beta}$ and $C_\rho^{\alpha \beta}$ are constant tensors. We have also assumed that dilaton field $\Phi$ is constant. This means that factor $e^\Phi$ is included in constants $f^{\alpha \beta}$ and $C_\rho^{\alpha \beta}$.
	This choice of background fields is accompanied with following condition
	\begin{align} \label{constraint}
		\gamma^\mu_{\alpha \beta} C_\mu^{\beta \gamma} = 0, \quad \gamma^\mu_{\alpha \beta} C_\mu^{\gamma \beta} = 0.
	\end{align}

	String propagates in superspace spanned by bosonic coordinates $x^\mu$ $(\mu = 0,1,...,9)$ and fermionic ones $\theta^\alpha$, $\bar{\theta}^\alpha$ with 16 independent real components each. Fermionic coordinates are accompanied by their canonically conjugated momenta  $\pi_\alpha$ and $\bar{\pi}_\alpha$. Both fermionic coordiantes and their momenta are given as Majorana-Wejl spinors. World sheet $\Sigma$ that string sweeps in this superspace is parameterized by $\xi^m$ $  (\xi^0 = \tau, \xi^1 = \sigma)$. By combining these parameters we can define light-cone parametrization $\xi^\pm = \frac{1}{2}(\tau \pm \sigma)$ and light-cone partial derivatives $\partial_\pm = \partial_\tau \pm \partial_\sigma$.
	
	Inserting all these assumptions into action (\ref{eq:dejstvo}) and integrating out fermionic momenta, we are left with following expression
	
	\begin{align} \label{Action final S}
		S       =    
		k    \int_{\Sigma}    d^2    \xi      
		\left[     \Pi_{ +\mu \nu }   \partial_+     x^\mu    \partial_-   x^\nu
		+     \frac{1}{2}      (  \partial_+  \bar{ \theta }^\alpha  +   \partial_+   x^\mu   \bar{ \Psi }_\mu^\alpha     )
		\left( F^{-1}  (x)   \right)_{ \alpha\beta }
		(    \partial_-    \theta^\beta      + \Psi_\nu^\beta      \partial_-  x^\nu         )
		\right] ,  
	\end{align}
	where we have introduced following tensors
	\begin{equation}
		\Pi_{ \pm \mu \nu } = B_{\mu \nu} \pm \frac{1}{2} G_{\mu \nu},
	\end{equation}
	\begin{equation}
		F^{\alpha\beta}(x)=f^{\alpha\beta}+C_\mu^{\alpha\beta} x^\mu\, ,\quad (F^{-1}(x))_{\alpha\beta}=( f^{ -1 } )_{ \alpha\beta }  - ( f^{ -1 } )_{ \alpha\alpha_1 }   C_\rho^{\alpha_1 \beta_1}    x^\rho   ( f^{ -1 } )_{ \beta_1\beta }\, .
	\end{equation}
	
	To obtain meaningful T-dual transformation laws we need to assume that $x^\mu$ dependent part of tensor $(F^{-1}(x))_{\alpha\beta}$ is antisymmetric and infinitesimal.
	This additional assumption does not infringe on constraint (\ref{constraint}), \cite{Vertex operators}.
	
	Having obtained one of relevant actions , we will now focus on bosonic T-dualization of (\ref{Action final S}) to obtain our second action of interest.
	
	\subsection{Bosonic T-dualization}
	
	Bosonic T-dualization of action (\ref*{Action final S}) is given in detail in \cite{our paper 3}. Here we will only summarize the most important results.
	
	One way to obtain T-duality is by Busher procedure. This procedure is based on localization of translation symmetry. When we localize symmetry we replace all partial derivatives with covariant ones, while in cases where background fields depend on coordinates we also need to introduce invariant coordinate. Invariant coordinate is non-local addition to action and it is the sole reason for emergence of non-commutative behavior in closed strings. Introduction of covariant derivatives and invariant coordinates produces additional gauge fields in action, which in turn add new degrees of freedom to the theory. T-dual and original theory represent same physical system and we expect that those two theories carry exact same degrees of freedom. Because of this, we remove all newly introduced degrees of freedom with Lagrange multipliers. By utilizing gauge freedom of action we can fix bosonic coordinates to be some constant, in essence removing them from action. This gauge fixed action is only a function of gauge fields and Lagrange multipliers. Finding equation of motion for Lagrange multipliers and inserting them into action we can restore original action. On the other hand, finding equations of motion for gauge fields and inserting them into action we obtain T-dual action.
	
	Action (\ref{Action final S}), due to antisymmetric part of tensor $F^{-1}_{\alpha \beta} (x)$ is invariant under global translations of bosonic coordinates. Following steps of Busher procedure, described in preceding paragraph, we obtain following T-dual action
	
	\begin{align}\label{bosonic T dual action}
		\begin{gathered}
			{}^b S = \frac{k}{2} \int_{\Sigma} d^2\xi \Big[ \frac{1}{2} \bar{\Theta}^{\mu\nu} \partial_+y_\mu \partial_- y_\nu +\partial_+ \bar{ \theta }^\alpha \left( {}^b F^{-1} \text{\small $( V^{(0)} )$}    \right)_{ \alpha\beta } \partial_- \theta^\beta
			\Big.\\
			\begin{aligned}
				\Big. +\partial_+y_\mu {}^b \bar{ \Psi }^{\mu \alpha}  \left( {}^b F^{-1}  \text{\small $( V^{(0)} )$}    \right)_{ \alpha\beta } \partial_- \theta^\beta
				+ \partial_+ \bar{ \theta }^\alpha \left( {}^b F^{-1}  \text{\small $( V^{(0)} )$}    \right)_{ \alpha\beta } {}^b \Psi^{\nu \beta} \partial_- y_\nu \Big].
			\end{aligned}
		\end{gathered}
	\end{align}
	
	Here, $y_\mu$ is a dual coordinate, left superscript ${}^b$ denotes bosonic T-duality and $V^0$ represents following integral
	
	\begin{align}
		\begin{gathered}
			\Delta V^{(0) \rho}  =\\   
			%%%%%%%%%%%%%%%%%%%%%%%%%%%%%%%%%%%%%%%%%%%%%%%%%%%%%%%%%%%%%%%%
			=\frac{1}{2} \int_{P} d \xi^+ \breve{\Theta}_-^{\rho_1 \rho} \left[ \partial_+ y_{\rho_1}     -  \partial_+ \bar{ \theta } ^\alpha ( f^{- 1 } )_{ \alpha\beta } \Psi_{\rho_1}^\beta
			\right]
			%%%%%%%%%%%%%%%%%%%%%%%%%%%%%%%%%%%%%%%%%%%%%%%%%%%%%%%%%%%%%%%
			- \frac{1}{2} \int_{P} d \xi^- \breve{\Theta}_-^{\rho \rho_1} 
			\left[    \partial_- y_{\rho_1}  +  \bar{ \Psi }_{\rho_1}^\alpha ( f^{- 1 } )_{ \alpha\beta } \partial_- \theta^\beta
			\right] .
		\end{gathered}
	\end{align}
	
	T-dual tensors that appear in action have following interpretation: $\bar{\Theta}_-^{\mu \nu}$ is inverse tensor of $\bar{\Pi}_{+\mu\nu} = \Pi_{ +\mu \nu } + \frac{1}{2} \bar{ \Psi }_\mu^\alpha \left( F^{-1}  ( x)   \right)_{ \alpha\beta }\Psi_\nu^\beta = \breve{\Pi}_{+\mu \nu } - \frac{1}{2}  \bar{ \Psi }_\mu^\alpha ( f^{ -1 } )_{ \alpha\alpha_1 }   C_\rho^{\alpha_1 \beta_1}    x^\rho   ( f^{ -1 } )_{ \beta_1\beta } \Psi_\nu^\beta $, defined as

	\begin{equation}
		\bar{\Theta}_-^{\mu\nu} \bar{\Pi}_{+\nu\rho} = \delta^\mu_\rho,
	\end{equation}
	where
	\begin{gather}
		\bar{\Theta}_-^{\mu\nu} = \breve{\Theta}_-^{\mu \nu}  + \frac{1}{2} \breve{\Theta}_-^{\mu \mu_1} \bar{ \Psi }_{\mu_1}^\alpha (f^{-1})_{\alpha \alpha_1} C^{\alpha_1 \beta_1}_\rho V^{(0)\rho} (f^{-1})_{\beta_1\beta} \Psi_{\nu_1}^{\beta_1}  \breve{\Theta}_-^{\nu_1 \nu} ,\\
		%%%%%%%%%%%%%%%%%%%%%%%%%%%%%%%%%%%%%%%%%%%%%%%%%%%%%%%%%%%%%%%%%%%%%%%%%%%%%%%%%%%%%%%%%%%%%%%%%%%%%%%%%%%%%%%%%%%%%%%%%%%%%%%%%
		\breve{\Theta}_-^{\mu \nu} \breve{\Pi}_{+\nu \rho} = \delta^\mu_\rho,\qquad \breve{\Theta}_-^{\mu\nu} = \Theta_-^{\mu\nu} - \frac{1}{2} \Theta_-^{\mu\mu_1} \bar{ \Psi }_{\mu_1}^\alpha (\bar{f}^{-1})_{\alpha\beta} \Psi^\beta_{\nu_1} \Theta_-^{\nu_1\nu}\\
		%%%%%%%%%%%%%%%%%%%%%%%%%%%%%%%%%%%%%%%%%%%%%%%%%%%%%%%%%%%%%%%%%%%%%%%%%%%%%%%%%%%%%%%%%%%%%%%%%%%%%%%%%%%%%%%%%%%%%%%%%%%%%%%%%
		\bar{f}^{\alpha\beta} = f^{\alpha \beta} + \frac{1}{2} \Psi_\mu^\alpha \Theta_-^{\mu\nu} \bar{ \Psi }_\nu^\beta,\\
		%%%%%%%%%%%%%%%%%%%%%%%%%%%%%%%%%%%%%%%%%%%%%%%%%%%%%%%%%%%%%%%%%%%%%%%%%%%%%%%%%%%%%%%%%%%%%%%%%%%%%%%%%%%%%%%%%%%%%%%%%%%%%%%%%
		\Theta_-^{\mu\nu} \Pi_{ +\mu \rho } = \delta^\mu_\rho, \qquad \Theta_- = -4 (G_E^{-1} \Pi_- G^{-1})^{\mu\nu}\, .
	\end{gather}
	
	Tensor $\left( {}^b F^{-1}  \text{\small $( V^{(0)} )$}    \right)_{ \alpha\beta }$ is T-dual to $\left( F^{-1}  ( x)   \right)_{ \alpha\beta }$,
	
	\begin{equation}
		\left( {}^b F^{-1} \text{\small $( V^{(0)} )$}    \right)_{ \alpha\beta } = \left( F^{-1}  \text{\small $( V^{(0)} )$}   \right)_{ \alpha\beta } - \frac{1}{2} \left( F^{-1} \text{\small $( V^{(0)} )$}   \right)_{ \alpha\alpha_1 } \Psi_\mu^{\alpha_1} \bar{\Theta}_-^{\mu \nu} \bar{ \Psi }_\nu^{\beta_1} \left( F^{-1}  \text{\small $( V^{(0)} )$}    \right)_{ \beta_1\beta }.
	\end{equation}
	
	Finally, $ {}^b \bar{ \Psi }^{\mu \alpha} $ and ${}^b \Psi^{\nu \beta}$ are T-dual gravitino fields, given as
	
	\begin{align}
		\label{Background fields transformation 2}
		&\begin{gathered}
			\begin{aligned}
				\quad{}^b \bar{ \Psi }^{\mu \alpha} = \frac{1}{2} \bar{\Theta}^{\mu\nu} \bar{ \Psi }_\nu^\alpha + \frac{1}{4} \bar{\Theta}_-^{\mu \mu_1} \bar{ \Psi }_{\mu_1}^\beta \left( F^{-1}  \text{\small $( V^{(0)} )$}    \right)_{ \beta\beta_1 }
				\Psi_\nu^{\beta_1} \Theta_-^{\nu \nu_1} \bar{ \Psi }_{\nu_1}^\alpha = \frac{1}{2} \Theta_-^{\mu\nu} \bar{ \Psi }_\mu^\alpha,
			\end{aligned}
		\end{gathered}\\
		&\begin{gathered}
			\begin{aligned}
				\quad{}^b \Psi^{\nu \beta} = -\frac{1}{2} \Psi_\mu^\beta \bar{\Theta}_-^{\mu\nu} -\frac{1}{2} \Psi_\mu^\beta \Theta_-^{\mu \mu_1} \bar{ \Psi }_{\mu_1}^{\alpha}  \left( F^{-1}  \text{\small $( V^{(0)} )$}   \right)_{ \alpha\alpha_1 }\Psi_{\nu_1}^{\alpha_1} \bar{\Theta}_-^{\nu_1 \nu}= - \frac{1}{2}\Psi_\mu^\beta \Theta_-^{\mu\nu}.
			\end{aligned}
		\end{gathered}
	\end{align}

	Having obtained actions (\ref{Action final S}) and (\ref{bosonic T dual action}), we can now consider dualization along fermionic coordinates.
	
	\section{Fermionic T-duality}
	\setcounter{equation}{0}
	
	In this section the objectives are to find fermionic T-dual transformation laws and actions that have been T-dualized along fermionic coordinates for case where we performed bosonic T-duality and case where we have not.
	
	Bosonic T-duality relies on utilization of symmetries of action to produce T-dual action and T-dual transformation laws. This task is usually accomplished by utilizing Busher procedure \cite{T-duality procedure 1} \cite{T-duality procedure 2} \cite{Generalized Buscher procedure} \cite{Gereralized Buscher procedure 2} \cite{Generalized Buscher procedure 3}. The main idea of fermionic T-duality is essentially the same, we utilize isometries of fermionic coordinates to generate T-dual action and T-dual transformation laws\cite{Fermionic T-duality}\cite{Fermionic T-duality 2}\cite{Fermionic T-duality 3}\cite{Fermionic T-duality 4}. Just like in bosonic case, we localize translational symmetry by introducing covariant derivatives and, in cases where necessary, invariant coordinates. After this we introduce term that eliminates additional degrees of freedom and gauge fix existing symmetry. From this point on, finding equations of motion for gauge fields and inserting those equations of motion into gauge fixed action we obtain T-dual action. 
	
	Before proceeding with fermionic variant of Busher procedure, we can notice that our actions (\ref{Action final S}) and (\ref{bosonic T dual action}) do not posses terms proportional to $\partial_+ \theta^\alpha$ and $\partial_- \bar{ \theta }^\alpha$. This means that our fermionic coordinates have following local symmetry
	
	\begin{equation}
		\delta \theta^\alpha = \epsilon^\alpha(\sigma^+), \quad \delta \bar{ \theta }^\alpha = \bar{\epsilon}^\alpha (\sigma^-), \quad (\sigma^\pm = \tau \pm \sigma).
	\end{equation}
	
	We need to fix this symmetry before obtaining T-dual theory, one way to do this is through BRST formalism. This symmetry has following corresponding BRST transformations for fermionic fields
	
	\begin{equation}
		s \theta^\alpha = c^\alpha(\sigma^+), \qquad s \bar{ \theta }^\alpha = \bar{c}^\alpha (\sigma^-).
	\end{equation}
	
	Here $s$ is BRST nilpotent operator, $c^\alpha$ and $\bar{c}^\alpha$ represent ghost fields that correspond to gauge parameters $\epsilon^\alpha$ and $\bar{\epsilon}^\alpha$ respectively. In addition to ghost fields we also have following BRST transformations 
	
	\begin{equation}
		sC_\alpha = b_{+\alpha}, \quad s\bar{C}_\alpha = \bar{b}_{-\alpha}, \quad sb_{+\alpha} = 0, \quad s\bar{b}_{-\alpha} = 0.
	\end{equation}
	where $\bar{C}_\alpha$ and $C_\alpha$ are anti-ghosts, $b_{+\alpha}$ and $\bar{b}_{-\alpha}$ are Nakanishi-Lautrup auxiliary fields.
	
	Fixing of gauge symmetry is accomplished by introduction of gauge fermion, where we have decided to follow in the same choice as \cite{Fermionic T-duality procedure}
	\begin{equation}
		\Psi = \frac{k}{2} \int_{\Sigma} d^2\xi \Big[ \bar{C}_\alpha \Big( \partial_+ \theta^\alpha + \frac{1}{2}\alpha^{\alpha \beta} b_{+ \beta}  \Big) + \Big( \partial_- \bar{ \theta }^\alpha + \frac{1}{2} \bar{b}_{- \beta} \alpha^{\beta \alpha}  \Big) C_\alpha
		\Big],
	\end{equation}
	
	here $\alpha^{\alpha \beta}$ is arbitrary invertible matrix.

	Applying BRST transformation to gauge fermion we obtain  gauge fixed action and Fadeev-Popov action
	
	\begin{align}
		S_{gf} &= \frac{k}{2} \int_{\Sigma} d^2 \xi \Big[
		\bar{b}_{-\alpha} \partial_+ \theta^\alpha + \partial_- \bar{ \theta }^\alpha b_{+ \alpha}  + \bar{b}_\alpha \alpha^{\alpha \beta} b_{+\beta}
		\Big], \\
		S_{F-P} &= \frac{k}{2} \int_{\Sigma} d^2 \xi \Big[
		\bar{C}_\alpha \partial_+ c^\alpha + (\partial_- \bar{c}^\alpha) C_\alpha
		\Big].
	\end{align}

	Fadeev-Popov term contains only ghosts and anti-ghosts and it is decoupled from the actions (\ref{Action final S}) and (\ref{bosonic T dual action}). From this point on, this term will be ignored. Gauge fixing term contains auxiliary fields $\bar{b}_{- \alpha}$ and $b_{+\alpha}$ that can be removed with equations of motion
	
	\begin{equation}
		\bar{b}_{- \alpha} = - \partial_- \bar{ \theta }^\beta (\alpha^{-1})_{\beta \alpha}, \qquad b_{+ \alpha} = - (\alpha^{-1})_{\alpha \beta} \partial_+\theta^\beta,
	\end{equation}
	giving us
	
	\begin{equation} \label{gauge fixing term}
		S_{gf} = -\frac{k}{2} \int_{\Sigma} d^2 \xi \partial_-\bar{ \theta }^\alpha (\alpha^{-1})_{\alpha\beta} \partial_+\theta^\beta.
	\end{equation}
	Inserting guage fixing term into (\ref{Action final S}) and (\ref{bosonic T dual action}) gives us actions that can be dualized with Busher procedure.
	
	\subsection{Type II superstring - fermionic T-duality}
	
	Since both action (\ref{Action final S}) and gauge fixing term (\ref{gauge fixing term}) are trivially invariant to global translations of fermionic coordinates, we localize this translational symmetry by replacing partial derivatives with covariant ones
	
	\begin{alignat}{2}
		&\partial_\pm \theta^\alpha     &&\rightarrow       D_\pm \theta^\alpha = \partial_\pm \theta^\alpha + u_\pm^\alpha, \\ 
		%%%%%%%%%%%%%%%%%%%%%%%%%%%%%%%%%%%%%%%%%%%%%%%%%%%%%%%%%
		&\partial_\pm \bar{ \theta }^\alpha     &&\rightarrow       D_\pm \bar{ \theta }^\alpha = \partial_\pm \bar{ \theta }^\alpha + \bar{u}_\pm^\alpha.
	\end{alignat}
	
	New gauge fields $u_\pm^\alpha$ and $\bar{u}_\pm^\alpha$ introduce new degrees of freedom that are removed by addition of term
	
	\begin{equation}
		S_{add} = \frac{k}{2}    \int_{\Sigma}       d^2       \xi          \left[ \bar{z}_\alpha (\partial_+ u_-^\alpha - \partial_- u_+^\alpha)  + (\partial_+ \bar{u}_-^\alpha - \partial_-\bar{u}_+^\alpha)z_\alpha
		\right].
	\end{equation}
	
	Gauge freedom can be utilized to fix fermionic coordinates such that $\theta^\alpha = \theta_0^\alpha = const$ and $\bar{ \theta }^\alpha = \bar{ \theta }_0^\alpha = const$. This in turn reduces our covariant derivatives to
	
	\begin{equation}
		D_\pm \theta^\alpha    \rightarrow      u_\pm^\alpha, \qquad
		%%%%%%%%%%%%%%%%%%%%%%%%%%%%%%%%%%%%%%%%%%%%%%%%%%%%%%%%%
		D_\pm \bar{ \theta }^\alpha    \rightarrow    \bar{u}_\pm^\alpha.
	\end{equation}
	
	With all this in mind, we have following action
	
	\begin{align} \label{Gauge fixed S}
		\begin{gathered}
			S _{gf}      =    
			k    \int_{\Sigma}    d^2    \xi      
			\Big[ \Big.    \Pi_{ +\mu \nu }   \partial_+     x^\mu    \partial_-   x^\nu
			+       \frac{1}{2}    ( \bar{u}_+^\alpha  +   \partial_+   x^\mu   \bar{ \Psi }_\mu^\alpha     )
			\left( F^{-1}  (x)   \right)_{ \alpha\beta }
			(   u_-^\beta   + \Psi_\nu^\beta      \partial_-  x^\nu         )\\
			- \frac{1}{2} \bar{u}_-^\alpha (\alpha^{-1})_{\alpha\beta} u_+^\beta + \frac{1}{2}
			\bar{z}_\alpha (\partial_+ u_-^\alpha - \partial_- u_+^\alpha)  + \frac{1}{2} (\partial_+ \bar{u}_-^\alpha - \partial_-\bar{u}_+^\alpha)z_\alpha \Big.\Big].
		\end{gathered}
	\end{align}
	
	On one side we have equations of motion for Lagrange multipliers $\bar{\chi}_\alpha$ and $\chi_\alpha$
	
	\begin{align}
		\partial_+ u_-^\alpha - \partial_- u_+^\alpha = 0 \quad &\rightarrow \quad u_\pm^\alpha = \partial_\pm \theta^\alpha,\\
		\partial_+ \bar{u}_-^\alpha - \partial_-\bar{u}_+^\alpha = 0\quad &\rightarrow \quad \bar{u}_\pm^\alpha = \partial_\pm\bar{ \theta }^\alpha.
	\end{align}
	
	Inserting solutions for these equations into action (\ref{Gauge fixed S}) we obtain starting action plus gauge fixing term.
	
	Variation of action with respect to gauge fields produces following set of equations of motion
	
	\begin{align}
		u_-^\alpha &= -\left( F^{\alpha \beta} (x) \partial_- z_\beta + \Psi_\mu^\alpha \partial_- x^\mu  \right),\\
		u_+^\alpha &= -\alpha^{\alpha \beta} \partial_+ z_\beta,\\
		\bar{u}_+^\alpha &= \partial_+ \bar{z}_\beta F^{\beta\alpha} (x) - \partial_+ x^\mu \bar{ \Psi }_\mu^\alpha,\\
		\bar{u}_-^\alpha &= \partial_-\bar{z}_\beta \alpha^{ \beta\alpha}.
	\end{align}
	
	Utilizing these equations we can remove gauge fields from action, resulting in action that depends only on Lagrange multipliers and bosonic coordinates
	
	\begin{equation} \label{fermionic T-dual only}
		\begin{gathered}
			{}^fS =k\int_{\Sigma} d^2 \xi \Big[\Big. \Pi_{ +\mu \nu } \partial_+ x^\mu \partial_- x^\nu + \frac{1}{2}\partial_+ \bar{z}_\alpha \Psi_\mu^\alpha \partial_-x^\mu + \frac{1}{2}\partial_+\bar{z}_\alpha F^{\alpha \beta}(x) \partial_-z_\beta \\
			- \frac{1}{2}\partial_+ x^\mu \bar{\Psi}_\mu^\alpha \partial_-z_\alpha  - \frac{1}{2}\partial_-\bar{z}_\alpha \alpha^{\alpha\beta} \partial_+z_\beta \Big.\Big].
		\end{gathered}
	\end{equation}
	
	Just like in the bosonic case, we have that left superscript ${}^f$ denotes fermionic T-duality. From here we can deduce background fields of feriomionic T-dual theory 
	\begin{align}
		{}^f \bar{\Pi}_{+\mu\nu} &= \Pi_{ +\mu \nu },\\
		{}^f \left( F^{-1}  (x)   \right)^{ \alpha\beta } &= F^{\alpha \beta} (x),\\
		{}^f \bar{ \Psi }_{\mu\beta}\  {}^f \left( F^{-1}  (x)   \right)^{ \beta\alpha} = - \bar{ \Psi }_\mu^\alpha \quad&\rightarrow\quad  {}^f \bar{ \Psi }_{\mu\beta} = -\bar{ \Psi }_\mu^\alpha \left( F^{-1}  (x)   \right)_{ \alpha\beta }, \\ 
		{}^f \left( F^{-1}  (x)   \right)^{ \alpha\beta} {}^f\Psi_{\mu\beta} = \Psi_\mu^\alpha \quad&\rightarrow\quad {}^f\Psi_{\mu\beta} = \left( F^{-1}  (x)   \right)_{ \beta \alpha}\Psi_\mu^\alpha.
	\end{align}
	
	Unlike bosonic case, fermionic T-dual theory is local. This can be attributed to the fact that background fields do not depend on fermionic coordinates. This in turn means that theory is geometric and we should not expect emergence of non-commutative phenomena.
	
	\subsection{Type II superstring - full T-duality}
	
	To obtain fully dualized theory we start with action that is already T-dualized along bosonic coordinates (\ref{bosonic T dual action}). Procedure for fermionic T-duality is mostly the same as described before. Only difference comes from the fact that bosonic T-duality introduced non-local term  $V^{0}$ which depends on $\theta^\alpha$ and $\bar{ \theta }^\alpha$ and now we need to introduce invariant fermionic coordinate in order for action to be invariant to local shift symmetry 
	
	%In this case procedure for fermionic T-duality is mostly the same as in previous section. We start by addition of gauge fixing term, localizing translational symmetry by replacing partial derivatives with covariant ones and adding Lagrange multipliers that eliminate additional degrees of freedom. Even thou we did not have background fields that depend on fermionic coordinates, by performing bosonic T-duality we introduced non-local term $V^{0}$ which depends on $\theta^\alpha$ and $\bar{ \theta }^\alpha$. This term makes it necessary to also introduce invariant coordinate in order to perform fermionic T-duality.
	
	\begin{align}
		D_\pm \theta^\alpha &= \partial_\pm \theta^\alpha + u_\pm^\alpha,\\
		D_\pm \bar{ \theta }^\alpha &= \partial_\pm \bar{ \theta }^\alpha + \bar{u}_\pm^\alpha,\\
		\theta^\alpha_{inv} = \int_{P} d \xi^m D_m \theta^\alpha &=  \int_{P} d \xi^m (\partial_m \theta^\alpha + u_m^\alpha) = \Delta \theta^\alpha + \Delta U^\alpha, \\
		\bar{ \theta }^\alpha_{inv} = \int_{P} d\xi^m D_m \bar{ \theta }^\alpha &= \int_{P} d \xi^m (\partial_m \bar{ \theta }^\alpha + \bar{u}_m^\alpha) = \Delta \bar{ \theta }^\alpha + \Delta \bar{U}^\alpha.
	\end{align}

	Fixing gauge symmetry as before, setting fermionic coordinates to constants, we deduce following relations 
	
	\begin{align}
		D_\pm \theta^\alpha \rightarrow  u_\pm^\alpha, \qquad
		D_\pm \bar{ \theta }^\alpha  \rightarrow  \bar{u}_\pm^\alpha, \qquad
		\theta^\alpha_{inv} \rightarrow \Delta U^\alpha, \qquad \bar{ \theta }^\alpha_{inv} \rightarrow \Delta \bar{U}^\alpha.
	\end{align}

	With these relations we obtain action that is only a function of gauge fields, lagrange multipliers and dual coordinates
	
	\begin{align}\label{gauge fixed action}
		\begin{gathered}
			{}^bS_{gf} = \frac{k}{2} \int_{\Sigma} d^2\xi \Big[ \frac{1}{2} \bar{\Theta}^{\mu\nu} \partial_+y_\mu \partial_- y_\nu + \bar{ u }_+^\alpha \left( {}^b F^{-1} \text{\small $( V^{(0)} )$}    \right)_{ \alpha\beta } u_-^\beta
			\Big.\\
			\begin{aligned}
				+\partial_+y_\mu {}^b \bar{ \Psi }^{\mu \alpha}  \left( {}^b F^{-1}  \text{\small $( V^{(0)} )$}    \right)_{ \alpha\beta } u_-^\beta
				+  \bar{ u }_+^\alpha \left( {}^b F^{-1}  \text{\small $( V^{(0)} )$}    \right)_{ \alpha\beta } {}^b \Psi^{\nu \beta} \partial_- y_\nu - \bar{ u }_-^\alpha (\alpha^{-1})_{\alpha \beta} u_+^\beta 
			\end{aligned}\\
			+\bar{z}_\alpha (\partial_+ u_-^\alpha - \partial_- u_+^\alpha)  + (\partial_+ \bar{u}_-^\alpha - \partial_-\bar{u}_+^\alpha)z_\alpha \Big.  \Big].
		\end{gathered}
	\end{align} 
	
	In order to simplify calculations we introduce following two substitutions
	
	\begin{equation}
		\left( {}^b F^{-1}  \text{\small $( V^{(0)} )$}    \right)_{ \alpha\beta } {}^b \Psi^{\nu \beta} \partial_- y_\nu + \partial_- z_\alpha = Z_{- \alpha},\qquad \partial_+y_\mu {}^b \bar{ \Psi }^{\mu \alpha}  \left( {}^b F^{-1}  \text{\small $( V^{(0)} )$}    \right)_{ \alpha\beta } - \partial_+ \bar{z}_\beta = \bar{Z}_{+ \beta}.
	\end{equation}
	
	Now, our action can be expressed as
	
	\begin{align}
		\begin{gathered}\label{action compact}
		{}^bS_{gf} = \frac{k}{2} \int_{\Sigma} d^2\xi \Big[ \frac{1}{2} \bar{\Theta}^{\mu\nu} \partial_+y_\mu \partial_- y_\nu + \bar{ u }_+^\alpha \left( {}^b F^{-1} \text{\small $( V^{(0)} )$}    \right)_{ \alpha\beta } u_-^\beta+ \bar{Z}_{+ \beta} u_-^\beta +\bar{u}_+^\alpha Z_{- \alpha}
			\Big.\\
			\begin{aligned}
				- \bar{ u }_-^\alpha (\alpha^{-1})_{\alpha \beta} u_+^\beta +\partial_- \bar{z}_\alpha u_+^\alpha - \bar{u}_-^\alpha \partial_+z_\alpha.
			\end{aligned}
		\end{gathered}
	\end{align}
	
	Similar to the first case, we can always revert to starting action by finding equations of motion for Lagrange multipliers and inserting their solutions into the action. In both cases equations of motion are the same so we take the freedom to omit them here.
	
	Equations of motion for gauge fields differ in this case. Since we have that $V^{(0)}$ depends on fermionic coordinates, equations of motion  have additional term that depends on invariant coordinate.

	\begin{equation} \label{tran law 1}
		u_+^\alpha = -(\alpha)^{\alpha \beta} \partial_+ z_\beta, \qquad \bar{ u }_-^\beta = \partial_- \bar{z}_\alpha (\alpha)^{\alpha\beta},
	\end{equation}
	\begin{align} \label{tran law 2}
		\bar{u}_+^\alpha = - \bar{ Z}_{+ \beta} \ {}^b F^{\beta\alpha}\text{\small $( V^{(0)} )$} - \beta_\nu^-\text{\small $( V^{(0)},U^{(0)} )$}\  {}^b\bar{ \Psi }^{\nu \alpha}, \\ \label{tran law 3}
		u_-^\beta = - {}^b F^{\beta\alpha}\text{\small $( V^{(0)} )$} Z_{- \alpha} - \beta_\mu^+\text{\small $( V^{(0)},U^{(0)} )$} \ {}^b \Psi^{\mu \beta}.
	\end{align}
	
	The beta functions, $\beta_\mu^\pm\text{\small $( V^{(0)},U^{(0)} )$}$, are obtained by varying $V^{(0)}$ (see \cite{our paper 3} for more details). They are given as
	
	\begin{align}\label{beta 1}
		\beta^\pm_{\mu}\text{\small $( V^{(0)},U^{(0)} )$} = & \mp\frac{1}{8}
		\partial_\mp\big[   \bar{ U }^\alpha      +     V^{\nu_1}     \bar{ \Psi }_{\nu_1} ^\alpha    \big]  
		( f^{ -1 } )_{ \alpha\alpha_1 }     C_\mu^{\alpha_1 \beta_1}      ( f^{ -1 } )_{ \beta_1\beta }
		\big[    U^\beta    + \Psi_{\nu_2} ^\beta  V^{\nu_2}        \big]\nonumber
		\\
		&\pm\frac{1}{8}
		\big[ \bar{ U }^\alpha     +     V^{\nu_1}     \bar{ \Psi }_{\nu_1} ^\alpha    \big]  
		( f^{ -1 } )_{ \alpha\alpha_1 }     C_\mu^{\alpha_1 \beta_1}      ( f^{ -1 } )_{ \beta_1\beta } 
		\partial_\mp\big[  U^\beta   + \Psi_{\nu_2} ^\beta  V^{\nu_2}       \big].
	\end{align}
	
	Inserting equations of motion for gauge fields into action (\ref{action compact}) and keeping only terms linear with respect to $C_\mu^{\alpha\beta}$, we obtain fully dualized action
	
	\begin{align}
		{}^{bf}S = \frac{k}{2} \int_{\Sigma} d^2\xi \Big[
		\frac{1}{2} \bar{\Theta}_-^{\mu\nu} \partial_+y_\mu \partial_-y_\nu - \bar{ Z}_{+\alpha} {}^b F^{\alpha\beta}\text{\small $( V^{(0)} )$} Z_{- \beta} - \partial_-\bar{z}_\alpha (\alpha)^{\alpha\beta} \partial_+z_\beta
		\Big].
	\end{align}
	
	Expanded, we have
	
	\begin{align}
		\begin{gathered} \label{full action dual}
			{}^{bf}S =k \int_{\Sigma} d^2\xi \Big[\Big.\frac{1}{4} \Theta_-^{\mu\nu} \partial_+y_\mu \partial_-y_\nu - \frac{1}{4}\partial_+y_\mu \Theta_-^{\mu\nu} \bar{ \Psi }_\nu^\alpha \partial_- z_\alpha    -\frac{1}{4}\partial_+\bar{z}_\alpha \Psi_\mu^\alpha \Theta_-^{\mu\nu} \partial_-y_\nu\\
			\begin{aligned}
				 +\frac{1}{2}\partial_+ \bar{z}_\alpha {}^b F^{\alpha\beta}\text{\small $( V^{(0)} )$} \partial_- z_\beta - \frac{1}{2} \partial_-\bar{z}_\alpha (\alpha)^{\alpha\beta} \partial_+z_\beta \Big.\Big].
			\end{aligned}
		\end{gathered}
	\end{align}
	
	From here, we can read background fields of T-dual theory
	%	{}^{bf} \left(F^{-1}   \text{\small $(x)$}    \right)^{ \alpha\beta } = {}^b F^{\alpha\beta}(x) \quad\rightarrow\quad {}^{bf} \left(F^{-1}   \text{\small $(x)$}    \right)^{ \alpha\beta }& = F^{\alpha \beta} \text{\small $(x)$}  + \frac{1}{2} \Psi_\mu^\alpha \Theta_-^{\mu \nu} \bar{ \Psi }_\nu^\beta,\\	

	\begin{equation}
		\begin{gathered}
		{}^{bf} \bar{\Pi}_+^{\mu\nu}= \frac{1}{4} \bar{\Theta}_-^{\mu \nu} - \frac{1}{2}{}^b \bar{ \Psi }^{\mu \alpha}\left( {}^b F^{-1}  \text{\small $( V^{(0)} )$}    \right)_{ \alpha\beta } {}^b \Psi^{\nu \beta} = \Theta_-^{\mu\nu},\\
		{}^{bf} \left(F^{-1}   \text{\small $(x)$}    \right)^{ \alpha\beta } = {}^b F^{\alpha\beta}(x)  = F^{\alpha \beta} \text{\small $(x)$}  + \frac{1}{2} \Psi_\mu^\alpha \Theta_-^{\mu \nu} \bar{ \Psi }_\nu^\beta,\\
		\begin{aligned}
			{}^{bf} \bar{ \Psi }^\mu_\alpha\  {}^{bf} \left(F^{-1}   \text{\small $(x)$}    \right)^{ \alpha\beta } = {}^b \bar{ \Psi }^{\mu\beta} =\frac{1}{2}\Theta_-^{\mu\nu} \bar{ \Psi }_\nu^\beta \quad&\rightarrow\quad {}^{bf} \bar{ \Psi }^\mu_\alpha = \frac{1}{2} \bar{\Theta}_-^{\mu\nu} \bar{ \Psi }_\nu^\beta \left( F^{-1}  \text{\small $(x )$}    \right)_{\beta \alpha },\\
			{}^{bf} \left(F^{-1}   \text{\small $(x)$}    \right)^{ \alpha\beta }\  {}^{bf}\Psi^\nu_\beta = {}^b \Psi^{\nu\alpha} = - \frac{1}{2} \Psi_\mu^\alpha \Theta_-^{\mu\nu} \quad&\rightarrow\quad {}^{bf}\Psi^\nu_\beta = -\frac{1}{2} \left( F^{-1}  \text{\small $(x )$}    \right)_{\beta \alpha } \Psi_\mu^\alpha \bar{\Theta}_-^{\mu\nu}.
		\end{aligned}
		\end{gathered}
	\end{equation}
	
	Comparing background fields in different stages of T-dualization we notice that both fermionic T-duality and bosonic T-duality affect all field, where all T-dual theories now have coordinate dependent fields. It should also be noted that non-commutative relations in theory emerge only after performing bosonic T-duality. Fermionic T-dual coordinates are always only proportional to fermionic momenta therefore Poisson brackets between fermionic coordiantes always remain zero.

	\subsection{Bosonic T-duality of fermionic T-dual theory}
	
	For completion sake, we will also T-dualize fermionic T-dual action (\ref{fermionic T-dual only}) along $x^\mu$ coordinates. In this specific case, where only RR field depends on bosonic coordinate, we expect that bosonic and ferionic T-dualities commute. Therefore, this section can be thought of as a check for calculations from previous section. 
	
	Bosonic T-duality is mostly the same as fermionic one \cite{Generalized Buscher procedure}\cite{Gereralized Buscher procedure 2}\cite{Generalized Buscher procedure 3}\cite{our paper 3}, where only difference is the lack of introduction of Fadeev-Popov and gauge fixing actions. We again start by localizing translational symmetry, inserting Lagrange multipliers and fixing gauge fields.  This produces following auxiliary action
	
		\begin{eqnarray}\label{eq:auxact}
		{}^f	S_{aux}&=&\kappa \int d^2\xi \Big[ v^\mu_+ \Pi_{+\mu\nu} v^\nu_- +\frac{1}{2}\partial_+ \bar z_\alpha f^{\alpha\beta} \partial_- z_\beta + \frac{1}{2}\partial_+ \bar z_\alpha C^{\alpha\beta}_\mu \partial_- z_\beta \Delta V^\mu  \Big. \nonumber \\
		\Big. &+&\frac{1}{2}\partial_+ \bar z_\alpha \Psi^\alpha_\mu v_-^\mu -\frac{1}{2}v^\mu_+ \bar\Psi^\alpha_\mu \partial_- z_\alpha  -\frac{1}{2} \partial_- \bar{z}_\alpha \alpha^{\alpha\beta} \partial_+z_\beta  + \frac{1}{2} y_\mu (\partial_+ v_-^\mu-\partial_- v^\mu_+) \Big].
	\end{eqnarray}
	
	Introducing the variables
	\begin{equation}
		Y_{+\mu}=\partial_+ y_\mu-\partial_+ \bar z_\alpha \Psi^\alpha_\mu\, ,\quad Y_{-\mu}=\partial_- y_\mu-\bar\Psi^\alpha_\mu \partial_- z_\alpha\, ,
	\end{equation}
	the action (\ref{eq:auxact}) gets much simpler form
	\begin{eqnarray}\label{eq:auxdej}
		{}^f S_{aux}&=&\kappa \int d^2\xi \Big[ v^\mu_+ \Pi_{+\mu\nu} v^\nu_- +\frac{1}{2}\partial_+ \bar z_\alpha f^{\alpha\beta} \partial_- z_\beta+\frac{1}{2}\partial_+ \bar z_\alpha C^{\alpha\beta}_\mu \partial_- z_\beta \Delta V^\mu  \Big. \nonumber \\
		\Big. &-& \frac{1}{2}Y_{+\mu}v^\mu_-+\frac{1}{2}v_+^\mu Y_{-\mu}\Big]\, .
	\end{eqnarray}
	Varying the above action with respect to gauge fields $v^\mu_+$ and $v^\mu_-$, we get, respectively,
	\begin{equation}\label{eq:v1}
		\Pi_{+\mu\nu} v^\nu_-=-\left(\frac{1}{2}Y_{-\mu}+\beta_{+\mu}(V)\right)\, , 
	\end{equation}
	\begin{equation}\label{eq:v2}
		v^\nu_+ \Pi_{+\nu\mu}=\frac{1}{2}Y_{+\mu}-\beta_{-\mu}(V)\, ,
	\end{equation}
	where $\beta_{\pm \mu}$ are the beta functions obtained from coordinate dependent term in the action
	\begin{equation}
		\beta_{\pm \mu}=\mp \frac{1}{8}\left(\bar z_\alpha C^{\alpha\beta}_\mu \partial_\mp z_\beta-\partial_\mp \bar z_\alpha C^{\alpha\beta}_\mu z_\beta\right).
	\end{equation}
	Inserting (\ref{eq:v1}) and (\ref{eq:v2}) into the auxiliary action (\ref{eq:auxdej}), keeping the terms linear in $C^{\alpha\beta}_\mu$, we obtain fully T-dualized action (first fermionic, then bosonic T-dualization)
	\begin{equation}
		{}^{fb} S=\kappa \int d^2\xi \left[\frac{1}{2}\partial_+ \bar z_\alpha F^{\alpha\beta}(\Delta V)\partial_- z_\beta+\frac{1}{4}Y_{+\mu}(\Pi_+^{-1})^{\mu\nu}Y_{-\nu}\right]\, .
	\end{equation}
	Expanding above action we prove that it is identical to one given in (\ref{full action dual}).
	
	\section{Few notes on non-commutativity}
	\setcounter{equation}{0}
	
	In paper \cite{our paper 3} it has been shown that bosonic T-duality produces non-commutative relations between bosonic T-dual coordinates. With this in mind, following question naturally arises: can we expect emergence of same behavior for fermionic coordinates after fermionic T-dualization? To get the answer for this question we have to express fermionic T-dual coordinates as some combination of starting coordinates and their momenta and connect T-dual Poisson brackets with Poisson brackets of original theory. Original theory is geometric theory with regular Poisson structure
	
	\begin{align}
		\{  x^\mu (\sigma) , \pi_\nu (\bar{\sigma})   \} = \delta^\mu_\nu \delta(\sigma - \bar{\sigma}), \quad \{ \theta^\alpha (\sigma), \pi_\beta (\bar{\sigma})  \} =  \{  \bar{ \theta }^\alpha (\sigma), \bar{\pi}_\beta (\bar{\sigma}) \} = \delta^\alpha_\beta \delta(\sigma- \bar{\sigma}),
	\end{align}
	where all other Poisson brackets vanish.

	We start with case that has only been T-dualized along fermionic coordinates. To find how T-dual coordinates depend on starting ones and their momenta we can begin by finding fermionic momenta of starting theory. It is useful to remember  that starting theory did not posses terms that are proportional to $\partial_+ \theta^\alpha$ and $\partial_- \bar{ \theta }^\alpha$ and that this symmetry was fixed with BRST formalism. Addition of gauge fixing term introduced modification to momenta of starting theory and to obtain correct non-commutative relations we should be working with theories that have gauge fixing term in them.
	With this in mind, it is easy to find fermionic momentum of original theory (\ref{Gauge fixed S})
	
	\begin{align}
		\pi_\beta &= -\frac{k}{2} \Big[   (\partial_+ \bar{ \theta }^\alpha + \partial_+ x^\mu \bar{ \Psi }_\mu^\alpha ) \left(F^{-1}   \text{\small $(x)$}    \right)_{ \alpha\beta } - \partial_- \bar{ \theta }^\alpha (\alpha^{-1})_{\alpha\beta}   \Big],\\
		\bar{\pi}_\alpha &= \frac{k}{2} \Big[    \left(F^{-1}   \text{\small $(x)$}    \right)_{ \alpha\beta } (\partial_- \theta^\beta  +   \Psi_\nu^\beta \partial_- x^\nu  )  - (\alpha^{-1})_{\alpha\beta} \partial_+ \theta^\beta   \Big].
	\end{align}
	
	Since we want to obtain Poisson brackets for equal $\tau$ we want to find $\sigma$ partial derivatives of dual coordinate
	
	\begin{align} \label{mom1}
		\partial_\sigma z_\alpha = \partial_+  z_\alpha - \partial_-  z_\alpha = \frac{2}{k} \bar{\pi}_\alpha,\\ \label{mom2}
		\partial_\sigma \bar{z}_\alpha = \partial_+ \bar{z}_\alpha - \partial_- \bar{z}_\alpha = -\frac{2}{k} \pi_\alpha.
	\end{align}
	
	Momenta of original theory commute with each other and with $x^\mu$ coordinates, therefore we deduce that there has been no change to geometric structure of this theory.
	
	For fully dualized theory, transformation laws (\ref{tran law 1}) (\ref{tran law 2}) (\ref{tran law 3}) all depend on dual bosonic coordinate however, when we insert transformation laws that connect original bosonic coordinates with T-dual ones (more details in \cite{our paper 3})
	
	\begin{align}\label{T^2 equations of motion 1}
		\partial_+y_\mu &= 2 \Big[ \partial_+ x^\nu \bar{ \Pi }_{+ \nu \mu}  + \beta^-_{\mu}(x) 
		\Big] + \partial_+ \bar{ \theta }^\alpha  \left(F^{-1} ( x)\right)_{\alpha \beta} \Psi_\mu^\beta, \\
		\label{T^2 equations of motion 2}
		\partial_-y_\nu &= - 2 \Big[ \bar{ \Pi }_{+ \mu \nu} \partial_-x^\nu +\beta^+_{\mu}(x)
		\Big]- \bar{ \Psi }_\nu^\alpha \left(F^{-1} (x)\right)_{\alpha \beta} \partial_- \theta^\beta \, ,
	\end{align}
	into transformation laws for fermionic coordinates (\ref{tran law 1}), (\ref{tran law 2}) and (\ref{tran law 3})  we again obtain relations (\ref{mom1}) and (\ref{mom2}).
	
	On a first glance it would seem that fermionic T-duality has not produced any new Poisson brackets, however this is not the case. While it is true that there are no modifications to Poisson brackets between fermions, we have new Poisson bracket structure between fermions and bosons. This can be seen from $\sigma$ derivative of bosonic T-dual coordinate
	
	\begin{equation}\label{sigma derivative of y}
		y'_\mu \cong\frac{\pi_\mu}{\kappa} +\beta^0_\mu(x), 
	\end{equation}
	where $\beta^0_\mu(x)$ is combination  $\beta^+_\mu(x) + \beta^-_\mu(x)$ given as
	
	\begin{align}
		\beta^0_{\mu}(x) = & \frac{1}{2}
		\partial_\sigma\big[   \bar{ \theta }^\alpha      +     x^{\nu_1}     \bar{ \Psi }_{\nu_1} ^\alpha    \big]  
		( f^{ -1 } )_{ \alpha\alpha_1 }     C_\mu^{\alpha_1 \beta_1}      ( f^{ -1 } )_{ \beta_1\beta }
		\big[    \theta^\beta    + \Psi_{\nu_2} ^\beta  x^{\nu_2}        \big] \nonumber
		\\
		-&\frac{1}{2}
		\big[ \bar{ \theta }^\alpha     +     x^{\nu_1}     \bar{ \Psi }_{\nu_1} ^\alpha    \big]  
		( f^{ -1 } )_{ \alpha\alpha_1 }     C_\mu^{\alpha_1 \beta_1}      ( f^{ -1 } )_{ \beta_1\beta } 
		\partial_\sigma\big[  \theta^\beta   + \Psi_{\nu_2} ^\beta  x^{\nu_2}       \big]\, .
	\end{align}
	
	Finding Poisson brackets between $\sigma$ derivatives of coordinates and integrating twice we obtain following relations
	
	\begin{align}
		\begin{gathered}
			\{ y_\mu(\sigma), \bar{z}_\beta  (\bar{\sigma})  \} =\\ \frac{1}{k} \Big[ \bar{ \theta }^\alpha (\sigma)     +     x^{\nu_1} (\sigma)    \bar{ \Psi }_{\nu_1} ^\alpha -2 \big(\bar{ \theta }^\alpha (\bar{\sigma})     +     x^{\nu_1} (\bar{\sigma})    \bar{ \Psi }_{\nu_1} ^\alpha\big)   \Big]  ( f^{ -1 } )_{ \alpha\alpha_1 }     C_\mu^{\alpha_1 \beta_1}      ( f^{ -1 } )_{ \beta_1\beta } H(\sigma - \bar{\sigma}), 
		\end{gathered}\\
		\begin{gathered}
			\{ y_\mu(\sigma), z_\alpha  (\bar{\sigma})  \} =\\
			\frac{1}{k}( f^{ -1 } )_{ \alpha\alpha_1 }     C_\mu^{\alpha_1 \beta_1}      ( f^{ -1 } )_{ \beta_1\beta } \Big[    
			\theta^\beta (\sigma)  + \Psi_{\nu_2} ^\beta  x^{\nu_2} (\sigma) - 2 \big(     \theta^\beta (\bar{\sigma})  + \Psi_{\nu_2} ^\beta  x^{\nu_2}  (\bar{\sigma})     \big) \Big]H(\sigma - \bar{\sigma}).
		\end{gathered}
	\end{align}

	\section{Conclusions}
	
	In this article we examined effects of fermionic T-dulity performed on action of type II superstring in pure spinor formalism. We carried out our investigation in two cases, one where we performed fermionic T-duality on previously non dualized action and in second case where we had action that was already dualized along bosonic coordinates. Starting (non dualized) action that we worked with described closed string that propagates in presence of Ramond-Ramond field with linear coordinate dependence. We made a decision to only consider dependence on bosonic coordinates, furthermore this dependence was tied to infinitesimal antisymmetric term $C_\mu^{\alpha\beta}$. Rest of the background fields were held constant. Terms in action that  were non-linearly dependent on fermionic coordinates were neglected. These choices were in accordance with consistency conditions for background fields and were made in order to keep calculations manageable.
	
	On the other hand, bosonic T-duality of starting action provided us with theory that was non-local. Unlike starting theory that only dependent on coordinates through RR field, this theory manifested coordinate dependence on all background fields. Furthermore, bosonic T-dual coordinates now exhibit non-commutative properties.
	
	Before we could start with T-dualization we noticed that both cases posses additional local symmetry which removed terms proportional to $\partial_+ \theta^\alpha$ and $\partial_- \bar{ \theta }^\alpha$. In order to obtain correct T-dual theory this symmetry was fixed through BRST formalism. In both cases procedure for obtaining fermionic T-duality was the same, we employed Busher T-dualizing procedure. Procedure is based on localization of translational symmetry where we replace partial derivatives with covariant ones. Introduction of covariant derivatives carries with itself new degrees of freedom in shape of gauge fields. By demanding that starting and T-dual theory give description of same physical system we inevitably demand for both theories to posses same degrees of freedom. Thus, all additional degrees of freedom must be removed with Lagrange multipliers. By utilizing gauge freedom we can also remove all instances of fermionic coordinates in action obtaining action that is only a function of gaguge fields and Lagrange multipliers. Finding equations of motion for gauge fields of this gauge fixed action and inserting their solutions into the action we obtain T-dual theory. 
	
	Carrying Busher procedure for fermionic coordinates of non dualized action we obtain local theory where all fields depend on bosonic coordinates. This theory is commutative, its Poisson brackets are identical to Poisson brackets of starting theory.
	
	Busher procedure in case of theory that has been dualizied along bosonic coordinates does not change coordinate dependence of the background fields. All fields are still dependent on both bosonic and fermionic coordinates and theory is still non-local. However, this theory posseses two additional non-trivial Poisson brackets. We have emergence of non-commutativity between bosonic and fermionic coordinates, where non-commutativity is proportional to infinitesimal constant $C_\mu^{\alpha\beta}$. 
	
	Same result is obtained even in case where we first perform fermionic and then bosonic T-duality. Commutativity between different dualities was expected since fully T-dual theory must be unique. Only distinction between different paths of T-dualization procedures can be noticed in intermediate theories, where most important change is transition of theory from being local to non-local.
	
	We suspect that it is possible to obtain T-dual theory that is fully non-commutative, theory that has non-commutativity even between fermionic coordinates, but we would need starting theory that has background fields that depend on both bosonic and fermionic coordinates.

\end{document}